\documentclass[aps,pre,twocolumn,showpacs,floatfix,preprintnumbers,amsmath,amssymb]{revtex4}
\usepackage{epsfig}
\usepackage{graphicx}
\usepackage{dcolumn}
\usepackage{bm}

\begin{document}
\title{Quantum spectrum as a time series : Fluctuation measures}
\author{M. S. Santhanam, Jayendra N. Bandyopadhyay and Dilip Angom}
\affiliation{Physical Research Laboratory,
Navrangpura, Ahmedabad 380 009, India.}

\begin{abstract}
The fluctuations in the quantum spectrum could be
treated like a time series. In this framework, we explore
the statistical self-similarity in the quantum spectrum using
the detrended fluctuation analysis (DFA) and random matrix theory (RMT).
We calculate the Hausdorff measure for the spectra of atoms and
Gaussian ensembles and study their self-affine properties.
We show that DFA is equivalent to $\Delta_3$ statistics of RMT, unifying
two different approaches.
We exploit this connection to obtain theoretical estimates for the Hausdorff measure.
\end{abstract}

\maketitle

 The fluctuations in physical systems carry important information about the
system. In the context of quantum systems, the spectral fluctuations reveal
whether the corresponding classical dynamics is regular or chaotic or a mixture of both \cite{boh1}.
It is well known that for the classically regular systems, the quantum eigenvalues
are uncorrelated while for the chaotic systems, the eigenvalues
tend to display a certain degree of correlation, which is dictated by the
symmetry properties of the system such as the presence or absence of
rotational and time reversal invariance.
Typically, the spectrum of high dimensional quantum systems like the
atoms, molecules and nuclei belong to the latter class.
This equivalence between the nature of classical dynamics and spectral fluctuations
in the corresponding quantized system
is generally through an analogy with random matrix ensembles \cite{boh1} and has
been verified in many simulations and experiments \cite{rmt}.

Recently a different approach has been suggested. It is possible to consider
suitably transformed eigenvalues of a quantum system as a time series.
Then, using time series analysis methods,
it was shown that the quantum systems display $f^{-\gamma}$ noise where
$1 \le \gamma \le 2$ depends on the classical dynamics in the system \cite{rel}.
In particular, for the chaotic systems the level
fluctuations display $1/f$ noise. So do the levels of
complex nuclei \cite{rel}.
In contrast, for the regular systems we get $1/f^2$ noise.
Hence, the spectral fluctuations can be characterized
without reference to random matrix models.

This is reminiscent of following types of time series.
In the Brownian motion or random walk time series $x_t, t=1,2,\dots,N$,
the successive increments of the series are uncorrelated. The variance
of this process is $\mbox{var}(x) \propto t^{2H}, (H=1/2)$, where $H$ is the
Hausdorff measure. In this case, the power 
spectrum gives rise to $1/f^2$ noise \cite{tur}.
Thus, {\em regular} classical systems with uncorrelated quantum levels
are analogous to a series of Brownian motion path. On the other hand, one could also imagine
a time series with $1/f$ noise. There are many mechanisms and processes that produce 
$1/f$ noise \cite{obf}. In a chaotic
system, the eigenvalues of the corresponding quantum system contains certain
degree of correlation and it displays $1/f$ noise. In this case, $H=0$ and corresponds
to variance being time-independent.
In general, for $0 < H < 1, \gamma = 2H + 1$. In terms of the exponent $H$,
our interest is in the
regular ($H=1/2$) and chaotic limits ($H=0$) for the fluctuations of the spectral `time series'.
This region $0 < H < 1/2$ corresponds to anti-persistent time series, i.e, the one which has
opposing trends at successive time steps. Visually an anti-persistent time series
presents a very rough profile as compared to $H=1/2$ series. In fact,
$H$ is also used to characterize the roughness of surfaces and profiles.

Such an anti-persistent series is also a self-affine fractal and displays statistical
self-similarity that scale differently in different directions.
Mathematically, it implies, $y(H t) \approx t^{-H} y(t)$
where the symbol $\approx$ represents statistical equality.
Note that the power spectrum is a measure
of the strength of frequencies in Fourier space but the fluctuation
function and the Hausdorff measure, taken together,
indicate the structure of the spectrum at various spectral scales. This
structure is related to the classical periodic orbits
of the system through semiclassical theories like
the Gutzwiller's trace formula \cite{gut}.
Hence, it is important to understand the self-affine properties of the
spectrum since it has implications at the level of classical
dynamics of the system.

The idea of characterizing fluctuations and its scaling is of interest in other
areas as well.  For instance,
the surface-height fluctuations in the surface growth processes reflect the
morphological changes that determine the physical and chemical processes
such as crystal growth, metal deposition etc. \cite{ro2}.
Fluctuations in the EEG series might tell us about the nature
of physiological process taking place \cite{eegdfa}. In all these cases, fluctuations
and how it scales provide an important clue to the physical process under study.
Hence, the results presented here have relevance in fields beyond quantum physics.

In this paper, we take the time series point of view for the quantum spectrum
and compute the exponent $H$ for levels of Lanthanide atoms (Sm, Pm and Nd) and 
RMT ensembles using DFA method \cite{dfap}. Note that the value
of $\gamma$ does not necessarily guarantee self-affine nature \cite{greis}.
We show that the DFA of order 1 is related to $\Delta_3$
statistics of RMT and exploit it to obtain theoretical fluctuation
functions and $H$.
We show that the time series of level sequences and a random time series both
with the same value of $H$ have the same fluctuation properties.

\begin{figure}
\includegraphics[height=7cm]{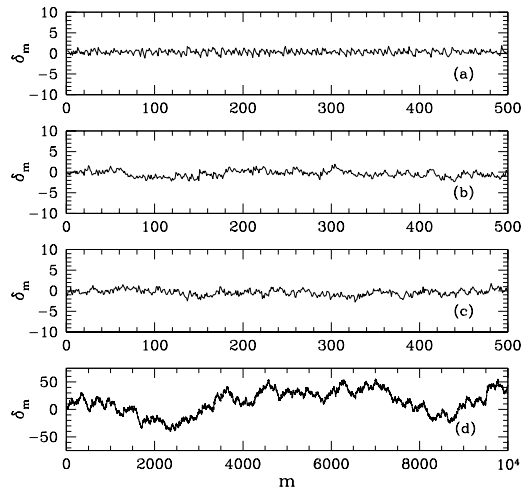}
\caption{$\delta_m-m$ curve for (a) Sm I atomic levels (b) GOE levels
(c) time series with $H=0$ and (d) GDE levels. Note that (a,b,c) have
qualitatively similar features as opposed GDE levels in (d).}
\label{dm-m}
\end{figure}

We denote a spectrum of discrete levels, belonging to either a random matrix
or a given atomic system, by $E_i, i=1,2,\dots,n+1$.
Then, its integrated level density, i.e, the
total number of levels below a given level, can be decomposed into an average part plus
an oscillatory contribution, $N(E) = N_{av}(E) + N_{osc}(E)$. Spectral unfolding is
effected by the transform, $\lambda_i = N_{av}(E_i)$, such that the mean level density
of the transformed sequence $\{ \lambda \}$ is unity. All further analysis is
carried out using $\{ \lambda \}$. For instance, the level spacing is, 
$d_i=\lambda_{i+1} - \lambda_i , i=1,2,\dots,n$. Main object of interest in this paper is the
fluctuations in the `time series' of levels given by,
\begin{equation}
\delta_m = \sum_{i=1}^{m} (d_i - \langle d \rangle ) \equiv - N_{osc}(E_{m+1})
; \; m=1,\dots,n.
\label{dn}
\end{equation}
Note that $\delta_m$ is formally analogous to a time series with the time index
denoted by $m$ \cite{rel}. The power spectrum $S(f)$
of $\delta_m$ displays a power law, i.e, $S(f) \sim f^{-\gamma}$, where
$1 \le \gamma \le 2$ \cite{rel}. In particular,
if the system is regular, $\gamma = 2$ and for Gaussian ensembles and the 
chaotic systems, $\gamma = 1$. Thus, $1/f$ noise should correspond to $H=0$.
A $H=0$ random time series should have a profile similar to
levels of Gaussian ensembles and atoms as seen in
Fig. \ref{dm-m}(a-c).

To obtain $\delta_m$, we needed to unfold the spectrum. This
puts the spectra from various systems on the same footing
amenable to direct comparison.
For the atomic systems,  unfolding was done using an empirical function
to fit the given levels. For the Gaussian ensembles the
Wigner's semi-circle law \cite{rmt} was used for unfolding.

First, we show the results from Lanthanide atoms. 
The atoms in this series exhibit complex configuration mixing and spectra.
Recently, a series of studies on the statistical properties of these levels from
RMT point of view were reported \cite{dk}. All the calculations for levels of Sm, Nd, Pm
were done using the {\sc grasp92} code and for details we refer to \cite{grasp}.
In Fig. \ref{dm-m}(a), we show $\delta_m$ for levels obtained from Sm. Note that the
profile is similar to levels from Gaussian Orthogonal Ensemble (GOE), shown in
Fig. \ref{dm-m}(b), and a random time series
with $H=0$ ( Fig. \ref{dm-m}(c) ). This is in contrast to the level profile of
Gaussian Diagonal Ensemble (GDE) ( Fig. \ref{dm-m}(d)) defined as an ensemble of
diagonal matrices whose elements
are Gaussian distributed random numbers.

To quantitatively obtain the exponent $H$, we compute it using the detrended fluctuation
analysis (DFA) technique, which has become a popular tool to study long range 
correlations in time series \cite{dfap}. DFA is based on the idea that if the 
time series contains
nonstationarities, then the variance of the fluctuations can be studied by successively
detrending using linear, quadratic, cubic ... higher order polynomials in a piecewise manner.
The entire time series is divided into segments of length $s$ and in each of them
detrending is applied and variance $F^2(s)$ about the trend is computed and averaged over
all the segments. This exercise is repeated for various values of $s$.
For a self-affine fractal, we have,
\begin{equation}
F(s) \propto s^{H}
\end{equation}
where $0 < H < 1$.
In Table \ref{table1}, we show the numerical estimates of
$H$ obtained for various atomic and Gaussian ensemble cases. The results tabulated alongside
represent average of $F(s)$ taken over an appropriate ensemble. For instance,
for the Sm levels, the ensemble consists of 6 sequences each with
650 levels. Similarly for Pm and Nd. In the case of Gaussian ensembles of RMT,
the results were obtained from ensemble of 50 level sequences each with 4600, 3900, 2100
levels respectively for Gaussian Orthogonal ensemble, Unitary ensemble (GUE) and Symplectic
ensemble (GSE). For the uncorrelated
levels from GDE, we get $H=0.508$ in good agreement with the expected value of 1/2.
For the atomic and Gaussian ensembles, the value
of $H$ is close to zero. In such cases standard DFA becomes inaccurate and
DFA is performed after integrating
$\delta_m$ once more so that the new exponent is $H_1 = H + 1$.
Table \ref{table1} shows that $H_1$ is close to unity for atomic and Gaussian ensembles.
Thus, the GDE levels are self-affine
fractal with $H=1/2$ but those of atomic and Gaussian ensembles $(H=0)$ do not fall
in this class.

\begin{table}
\caption{\label{table1} Numerical estimates of $H_1$ and $H$.}
\begin{ruledtabular}
\begin{tabular}{lcc}
System & Hausdorff measure ($H$) & $H_1$ \\
\hline
Sm  & 0.045 & 1.045 $\pm$ 0.031   \\
Pm  & -0.025 & 0.975 $\pm$ 0.024  \\
Nd  & -0.003 & 0.997  $\pm$ 0.013 \\
GOE & 0.0003 & 1.0003 $\pm$ 0.0083 \\
GUE & 0.0047 & 1.0047  $\pm$ 0.0089 \\
GSE & 0.0025 & 1.0025 $\pm$ 0.0125  \\
GDE & 0.5080 &  1.5080 $\pm$ 0.0044 \\
\end{tabular}
\end{ruledtabular}
\end{table}

\begin{figure}
\includegraphics[height=7cm]{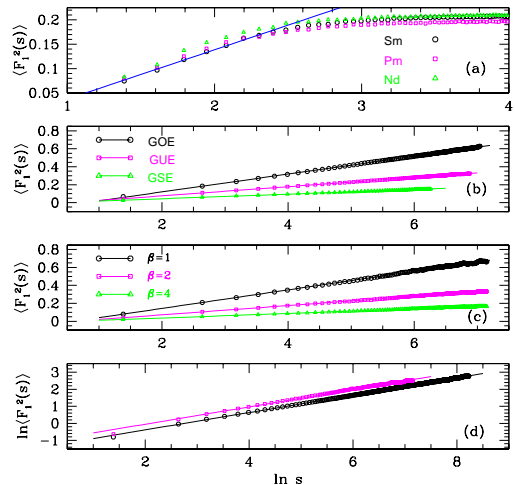}
\caption{(Color online) The fluctuation function $\langle {F_1}^2(s) \rangle$
in a semi-log plot for (a) atomic levels, (b) Gaussian ensembles
and (c) random time series with $H=0$ and power spectrum $1/\beta f$.
(d) shows $\ln \langle {F_1}^2(s) \rangle$ vs $\ln s$ for GDE and a random
walk time series. The slopes of best fit (solid) lines are quoted
in the text.}
\label{d3figs}
\end{figure}

In order to obtain analytical estimate for the value of $H$, we appeal to
the $\Delta_3(s)$ statistic widely studied in RMT to quantify rigidity of
spectrum, i.e, given a level, a rigid spectrum will not allow any flexibility
in the placement of the next level. $\Delta_3(s)$ is a measure of
this flexibility or conversely the rigidity. It is defined by,
\begin{equation}
\Delta_3(\bar{\lambda},s) = \frac{1}{2s} \min_{a,b} \int^{\bar{\lambda} + s}_{\bar{\lambda} - s} [n(\lambda) - a \lambda - b]^2 d\lambda
\label{d3}
\end{equation}
where $n(\lambda) = n_{av}(\lambda) + n_{osc}(\lambda) $ is the integrated
level density for the unfolded spectrum $\{ \lambda \} $ and
$a$ and $b$ are the parameters from linear regression and $\bar{\lambda}$ is any
reference level. Now, $\Delta_3(s)$ is an average over various choices of $\bar{\lambda}$.
Notice that
$ n(\lambda) = N(E) $ and due to unfolding procedure, $ n_{av}(\lambda) = \lambda $.
This implies that, $ n_{osc}(\lambda) = N_{osc}(E) $. Then, we could rewrite
Eq. \eqref{d3} as,
\begin{equation}
\Delta_3(\bar{\lambda},s) = \frac{1}{2s} \min_{a,b} \int^{\bar{\lambda} + s}_{\bar{\lambda} - s} [n_{osc}(\lambda) - (a-1) \lambda - b]^2 d\lambda
\label{d3mod}
\end{equation}
Now we discretize this equation to relate $\Delta_3(s)$ with the standard fluctuation function $F(s)$
obtained by DFA. Substituting from Eq. \eqref{dn} for $n_{osc}(\lambda)$, we 
get,
\begin{equation}
\begin{split}
\Delta_3(s) =& \frac{1}{M} \sum_{j=0}^{M} \frac{1}{2s} \sum_{m=j+1}^{j+2s} 
[ \delta_m - Y(\lambda_m) ]^2\\
=& \frac{1}{M} \sum_{j=0}^{M} F^2(s,j) = F^2(s)
\end{split}
\label{d3n}
\end{equation}
where $Y(\lambda_m)=(a^*-1)\lambda_m + b^*$ with best fit parameters $a^*$ and $b^*$ and
$j$ indexes $M$ different choices of $\bar{\lambda}$.
Since the DFA is applied with linear detrending on
segments of length $s$, then Eq. \eqref{d3n} boils down to $\Delta_3(s) = F_1^2(s)$,
and the subscript denotes DFA(1).
Then, ensemble averaging and writing $F_1(s)$ in terms of $\Delta_3(s)$, we have,
\begin{equation}
\langle F_1(s) \rangle =  \langle \Delta_3(s)^{1/2} \rangle
\label{dfatod3}
\end{equation}
This main result of the paper shows DFA(1) to be equivalent to $\Delta_3$ statistic.
From RMT, the averages for $\langle \Delta_3(s) \rangle $ are well known. For the
GDE, we have $\langle \Delta_3^{\mbox{GDE}}(s) \rangle = s/15$. Then, the result for uncorrelated set of
levels is,
\begin{equation}
\langle F_1(s) \rangle \sim s^{H} \sim \left( \frac{s}{15} \right)^{1/2}
\end{equation}
On a log-log plot we expect a slope of $H=1/2$. This is amply verified
in Fig. \ref{d3figs}(d) for GDE with slope 0.508. In our numerical simulation, GDE consists of
50 level sequences of 15000 levels each.

The numerical results for atomic levels and Gaussian ensembles in Table \ref{table1} show that
$H \approx 0 ~ (H_1 \approx 1)$. To make correspondence with $H_1$, we consider modified
$\Delta_3$ statistics given by, 
\begin{equation}
\Delta(\bar{\lambda},s) = \frac{1}{2s} \min_{a,b} \int^{\bar{\lambda} + s}_{\bar{\lambda} - s} [ \lambda~n(\lambda) - a \lambda - b]^2 d\lambda
\label{d3modif}
\end{equation}
which corresponds to integrating the $\delta_m$ once more before performing DFA on it. We obtain the
asymptotic random matrix average to be \cite{new},
\begin{equation}
\langle \Delta(s) \rangle \sim s^{2H_1} \propto s^{2}
\end{equation}
for all the three ensembles. Then, using Eq \ref{dfatod3},
$\langle F_1(s) \rangle \propto s$ and hence the theoretical slope
$H_1=1$ and corresponds to standard Hausdorff measure $H=0$.
The numerical values in last column of Table \ref{table1} confirm this RMT based result.
For Pm and Nd, note that the small negative values for $H$ are in the
error bar as seen by the error estimates provided.

Using the time series analogy, we explore the nature
of fluctuations at $H=0$ for Gaussian ensembles. This, in turn, throws further
light on the fluctuation properties of $H=0$ time series. From RMT we have,
$\langle \Delta_3(s) \rangle_{\beta} \approx \frac{\ln 2\pi s}{\beta \pi^2}$, where $\beta=1,2,4$ for
GOE, GUE and GSE respectively. Substituting this in Eq. \eqref{dfatod3},
the fluctuation function at $H=0$ is,
\begin{equation}
\langle {F_1}(s) \rangle \sim \left( \frac{\ln 2\pi s}{\beta \pi^2} \right)^{1/2}
\label{dd1}
\end{equation}
At $H=0$, the ensemble averaged fluctuations display logarithmic scaling with length of
level sequence $s$ considered. For convenience, we plot $\langle {F_1}^2(s) \rangle$
in a semi-log plot.  Then we expect slopes for level sequences from GOE, GUE and GSE to be  
$1/\beta \pi^2$. In Fig. \ref{d3figs}(a), we show results from atomic levels of
Sm, Pm and Nd. In all the three cases, the straight-line obtained shows that the 
DFA fluctuation function is in good agreement
with the theoretical result based on $\Delta_3(s)$ statistics.
Their slopes, 0.105, 0.104 and 0.109 for Sm, Pm and Nd
agree closely with $1/\pi^2$. The deviations for $\ln s > 2.2$ occurs due to
breakdown of universality in the presence of system specific information.
This is related to the period of the short periodic orbits in the atomic system \cite{berry}.
Thus, quantum spectrum of complex atoms can be characterized by $H=0$ and
display logarithmic scaling. Similar results are obtained for the Gaussian ensembles.
Fig. \ref{d3figs}(b) shows linear relationship in semi-log plot for levels from GOE, GUE and
GSE.
The measured best fit slopes 0.0995 $(\beta=1)$, 0.0511 $(\beta=2)$, 0.0257 $(\beta=4)$ closely approximate
the theoretical value $1/\beta\pi^2$. The DFA applied to spectral levels can
distinguish between the three ensembles. This is not possible with the power spectrum
method unless one measures the slope and the intercept together. Intercept measurement
is often difficult due to deviations from scaling for small frequencies.

What happens to an anti-persistent time series at $H=0$ ? We generate random
time series with $H=0$ and power spectrum $S(f) = 1/2\beta\pi^2 f$.
The technique is to generate Gaussian random numbers $u$ and take its Fourier transform 
to obtain $\tilde{u}$. Then, take the inverse Fourier transform of $\tilde{u} \sqrt{|S(f)|}$
to obtain a time series with the required properties \cite{rang}.
By varying $\beta$, we get a synthetic random time series
that will mimic level fluctuations
of GOE, GUE and GSE. The DFA applied to 10 member ensemble of
such time series (each of length 5000) is shown
in Fig. \ref{d3figs}(c). On a semi-log plot, they produce the expected linearity
and slopes 0.1035, 0.0517 and 0.0258 agree quite well with the theoretically expected
result $1/\beta \pi^2$. We conclude that an appropriate time series with $1/f$ noise
is similar to a level spectrum from random matrix ensembles and complex
atoms and both display fluctuations that scale in a logarithmic manner at $H=0$.
These results, viewed in conjunction with Bohigas {\it et. al.}  conjecture \cite{boh1},
mean that we can expect the spectrum of classically regular systems
to be characterized by $H=1/2$ and chaotic systems by $H=0$.

Even though we motivated our work using quantum systems, we emphasize that if
the spectrum is viewed as a time series, then we can apply our main
results to any classical system where self-affinity of the time series is relevant.
For instance, in studies of surface roughness,
growth models where surface morphology determines the properties
of the physical and chemical processes. Examples of surface growth
are the evolution of landscapes, vapor deposition, propagation
of flame fronts and growth of bacterial colonies etc. \cite{ro2}. 
In these cases, a
relevant quantity of interest is the correlation function for the surface heights
$h({\mathbf x})$,
$C({\mathbf r}) = \langle |h({\mathbf x}) - h({\mathbf x+r}) |^2 \rangle^{1/2}$.
In many cases, $C(r) \sim r^H$ over a large range of length scales, i.e,
many such growing surfaces are self-affine fractals.
In Ref. \cite{pal} a modified form for the correlation
function of rough surfaces is studied which gives logarithmic fluctuations, similar 
to Eq. \eqref{dd1} obtained above,
as $H \to 0$. This has been confirmed by numerical simulations of surfaces as well as
from experimental results of thin-film evaporation \cite{pal}. Another feature of
surface growth is the existence of roughening transition for certain critical
parameter values. At roughening transition, the exponent is $H=0$ and $C(r)$ displays 
logarithmic fluctuations as experimentally verified for Ag(115) sample \cite{ro1}.
We have also shown that $H=0$ corresponds to log fluctuations using random matrix methods.
Hence, taking the time series perspective and RMT could be beneficial for
various other problems.

To summarize, we treat the quantum spectrum like a time series.
In this picture, the uncorrelated spectral levels behave like
a random walk series with Hausdorff measure $H=1/2$ and is
a self-affine fractal.  For the spectrum of complex atoms
and Wigner-Dyson random matrix ensembles, $H=0$, an anti-persistent
time series.
We show that the DFA is equivalent to $\Delta_3$ statistics of RMT and we exploit
this connection to obtain theoretical estimates for $H$. This shows
that at $H=0$ the spectral fluctuations display logarithmic behavior, a feature
seen in many experimental and model simulations in surface growth studies.
Thus, $H=0$ is an unusual and interesting limit and will require further
investigation.

\thanks

We thank Prof. J. C. Parikh, Dr. P. Panigrahi, Prof. V. B. Sheorey and
P. Manimaran for useful discussions.

\end{document}